\documentclass[aps,prl,twocolumn,showpacs,preprintnumbers,superscriptaddress,dblfloatfix]{revtex4-1}
\usepackage{graphicx}   
\usepackage{dcolumn}    
\usepackage{bm}         
\usepackage{amssymb}    
\usepackage{setspace}
\usepackage{amsmath, amssymb, setspace}
\usepackage{array}
\usepackage{booktabs}
\usepackage{mathrsfs}
\usepackage{indentfirst}
\usepackage{slashed}
\usepackage{float}
\usepackage{lmodern}
\usepackage{hyperref}
\usepackage{xcolor}
\usepackage{multirow}
\usepackage{epstopdf}
\usepackage{ulem}
\usepackage{tabularx}
\usepackage[justification=raggedright]{caption}
\usepackage[flushleft]{threeparttable}
\usepackage{hyperref}
\usepackage{soul}

\begin{document}


\title{Positrons from Primordial Black Hole Microquasars and Gamma-ray Bursts}

\author{Volodymyr Takhistov}
\email[]{vtakhist@physics.ucla.edu}
\affiliation{Department of Physics and Astronomy, University of California, Los Angeles\\
Los Angeles, CA 90095, USA}

\date{\today}

\begin{abstract}
We propose several novel scenarios how  capture of small sublunar-mass primordial black holes (PBHs) by compact stars, white dwarfs or neutron stars, can lead to distinct short gamma-ray bursts (sGRBs) as well as microquasars (MQs).~In addition to providing new signatures, relativistic jets from these systems will accelerate positrons to high energies.~We find that if PBHs constitute a sizable fraction of DM, they can significantly contribute to the excess observed in the positron flux by the Pamela, the AMS-02 and the Fermi-LAT experiments. Our proposal combines the beneficial features of astrophysical sources and dark matter.
\end{abstract}

\maketitle

{\bf  Introduction  -- }
Primordial black holes (PBHs) can appear from early Universe dynamics and account for all or part of the dark matter (DM)~\cite{Zeldovich:1967,Hawking:1971ei,Carr:1974nx,GarciaBellido:1996qt,Khlopov:2008qy,Frampton:2010sw,Kawasaki:2012kn,Kawasaki:2016pql,Cotner:2016cvr,Carr:2016drx,Inomata:2016rbd,Inomata:2017okj,Georg:2017mqk}. Aside from being a theoretical curiosity, null search results for conventional DM particle candidates~\cite{Feng:2010gw,Bertone:2004pz} as well as the possible implications \cite{Bird:2016dcv,Nakamura:1997sm,Clesse:2015wea,Raidal:2017mfl,Eroshenko:2016hmn,Sasaki:2016jop,Clesse:2016ajp,Takhistov:2017bpt} for the newly-opened field of gravity-wave astronomy \cite{Abbott:2016blz,Abbott:2016nmj,Abbott:2017vtc} further elevate the interest in PBH-related studies.~Recent investigations have shown that PBHs could shed light on a variety of outstanding astronomical puzzles, such as the origin of $r$-process nucleosynthesis material \cite{Fuller:2017uyd}.

Observations by several experiments, including PAMELA~\cite{Adriani:2013uda}, Fermi-LAT~\cite{FermiLAT:2011ab} and AMS-02~\cite{Aguilar:2013qda}, have identified a rise in positron cosmic ray flux above $\sim10$ GeV. The origin of this phenomenon remains elusive.~A multitude of proposals have been put forward to address it, which can be generally grouped together as those based on astrophysical sources (e.g. pulsars~\cite{Hooper:2008kg,Yuksel:2008rf,Kawanaka:2009dk,Profumo:2008ms,Aharonian:1995zz,Malyshev:2009tw}, supernova remnants \cite{Kobayashi:2003kp,Shaviv:2009bu,Blasi:2009hv,Mertsch:2009ph,Biermann:2009qi}, microquasars \cite{Heinz:2002qj}), particle dark matter annihilations/decays \cite{ArkaniHamed:2008qn,Cirelli:2008pk,Chen:2008yi,Hooper:2008kv} as well as those based on cosmic ray propagation effects \cite{Blum:2013zsa,Delahaye:2007fr,Stawarz:2009ig}.

In this \textit{Letter} we discuss how PBHs interacting with compact stars can incite distinct gamma-ray burst (GRBs) and microquasar (MQs) sources, which can accelerate particles to high energies and contribute to the positron excess.~Heuristically, if a small PBH with sublunar mass of $10^{-16} M_{\odot} \lesssim M_{\rm PBH} \lesssim 10^{-7} M_{\odot}$ is captured by a compact star \cite{Capela:2013yf}, a white dwarf (WD) or a neutron star (NS), it will eventually consume the host and result in a stellar-mass BH.~The system's energy, released on dynamical time-scales, is sufficient to power a short GRB. In a different scheme, the resulting stellar-mass BH could steadily accrete matter if the considered star system is a binary, powering a microquasar jet. While a GRB explosion provides a singular energy injection, a microquasar jet is a continuous injection source.~Relativistic positrons, accelerated either through a burst or a continuous jet, will diffuse and are observable.

{\bf Black hole capture -- }
A small PBH can become gravitationally captured by a NS or a WD if it loses sufficient energy through dynamical friction and accretion as it passes through the star.~We briefly review the main capture ingredients, following \cite{Capela:2013yf, Takhistov:2017bpt}.~The full capture rate is given by $F = (\Omega_{\rm PBH} / \Omega_{\rm DM}) F_0$, where $\Omega_{\rm PBH}$ is the PBH contribution to the overall DM abundance $\Omega_{\rm DM}$.
The base Galactic capture rate $F_0$ is   
\begin{equation} \label{eq:f0calc}
F_0 = \sqrt{6 \pi} \dfrac{\rho_{\text{DM}}}{M_{\rm PBH}} \Big[\dfrac{R_{\rm NS} R_s}{\overline{v}(1 - R_s/R_{\rm NS})}\Big] \Big(1 - e^{- E_{\rm loss}/E_b}\Big)~,
\end{equation}
where $\rho_{\rm DM}$ is the DM density, $M_{\rm PBH}$ is the PBH mass, $\overline{v}$ is the DM velocity dispersion (assumed to have Maxwellian distribution), $E_b = M_{\rm PBH}\overline{v}^2/3$, $R_{\rm NS}$ is the radius of the NS with mass $M_{\rm NS}$ and Schwarzschild radius $R_{\rm S} = 2 G M_{\rm NS}$.~Convention $c = 1$ is used throughout. If the interaction energy loss $E_{\rm loss}$ exceeds kinetic energy of the PBH, then it will be captured.~The average energy loss for a NS is $E_{\rm loss} \simeq 58.8 \,G^2 M_{\rm PBH}^2 M_{\rm NS}/R_{\rm NS}^2$.~Throughout this work we consider a typical NS to have radius $R_{\rm NS} \sim 12$ km, mass $M_{\rm NS} \sim 1.5 M_{\odot}$ and spinning with a milli-second period $P \sim 1$ ms (i.e. a milli-second pulsar) at an angular velocity  $\Omega_{\rm NS} = 2 \pi/P$. In the case of WDs the star's mass is some-what lower, but the radius is significantly larger. This results in the WD capture rate being several orders below that of NSs. The total number of PBHs captured within time $t$ is $F t$.

After capture, PBH will settle inside the star and consume it through Bondi spherical accretion.~For a typical NS, the time for captured PBH to settle within is $t_{\rm set}^{\rm NS} \simeq  9.5 \times 10^3 (M_{\rm PBH}/10^{-11} M_{\odot})^{-3/2}$ yrs.~For a WD it is  
$t_{\rm set}^{\rm WD} \simeq  6.4\times10^6 (M_{\rm PBH}/10^{-11} M_{\odot})^{-3/2}$ yrs.~Once settled, the time for the black hole to consume the star is 
$t_{\rm con}^{\rm NS} \simeq 5.3 \times 10^{-3} (10^{-11} M_{\odot}/M_{\rm PBH})$ yrs for a NS and $t_{\rm con}^{\rm WD} \simeq 2.9 \times 10^2 (10^{-11} M_{\odot}/M_{\rm PBH})$ yrs for a WD, respectively.
If interaction time exceeds the timescales associated with the above processes, the system will effectively contain a stellar-mass BH once a PBH has been captured.~This outcome does not strongly depend on the star's equation of state (EoS).

We note, in passing, that a system with a BH inside a star is reminiscent of a Thorn-Zytkow object \cite{ThorneZytkow:1977}, a ``hybrid star" with a compact star residing within another star, which might form as a result of e.g. NS-WD merger~\cite{Paschalidis:2011ez}. 
 
{\bf Gamma-ray bursts -- }
Short GRBs~\cite{Berger:2013jza,Lee:2007js} are irregular electromagnetic pulses that last $\sim0.1-2$~s with a total $\gamma$-ray energy release of $\sim10^{48}-10^{50}$ erg. The prevalent progenitor scenario for their origin is binary compact object mergers, NS-NS~\cite{Eichler:1989ve} or BH-NS~\cite{Narayan:1992iy}.
Recent simulations have explicitly demonstrated that BH-NS and NS-NS
binaries can launch sGRB jets \cite{Paschalidis:2014qra, Ruiz:2016rai}.
~This picture is further supported by the recent observation of a NS-NS merger with accompanying electromagnetic emission \cite{TheLIGOScientific:2017qsa}.~Another possibility is an accretion-induced collapse of NSs~\cite{MacFadyen:2005xm,Dermer:2006pw,Vietri:1999kj}.~Long GRBs (timescale $>2$~s) are thought to originate from ``failed supernovae'' of massive stars (i.e. ``collapsars'') ~\cite{MacFadyen:1998vz}.~A common theme in the above scenarios is a resulting BH that is expected to be engulfed in debris forming a surrounding accretion disk/torus. The disk is rapidly accreted on dynamical timescales and the system releases energy as a GRB.  
As we argue, disk formation and the resulting BH-disk system could also be a likely consequence of PBH-induced NS implosions. Further, we outline several generic channels of sGRB production for this setup. 
We note that while GRBs originating from PBHs have been previously suggested in \cite{Cline:1992ps} and \cite{Derishev:1999vj}, these scenarios are strongly model-dependent and rely either on hypothetical ``explosions'' of evaporating PBHs at the end of their life or the existence of stable quark-star configurations from NS phase-transitions, respectively.

A sizable accretion disk can generically form during collapse of a uniformly rotating star \cite{Shapiro:2004zx}.~The size of the disk, however, strongly depends on the initial conditions of the system, such as the star's equation of state. The EoS can be modeled by a polytropic relation $P = K \rho^{1 + 1/n}$, where $P$ is the pressure, $K$ is a constant, $\rho$ is the density and $n$ is the polytropic index. For NSs,
recent observations of $\sim2 M_{\odot}$ pulsars J1614-2230 \cite{Demorest:2010bx} and J0348+0432 \cite{Antoniadis:2013pzd} favor a stiffer EoS, corresponding to a smaller polytropic index and a more uniform star density profile.~In \cite{Margalit:2015qza}, however, it was shown that for stiffer EoS formation of an accretion disk from a collapsing rotating massive NS is problematic. With a PBH inside, the system could behave differently.

\begin{figure}[tb]
\centering
\hspace{-2em}
\includegraphics[trim = 0.0mm .0mm 20mm 0mm, clip,width = 3.5in]{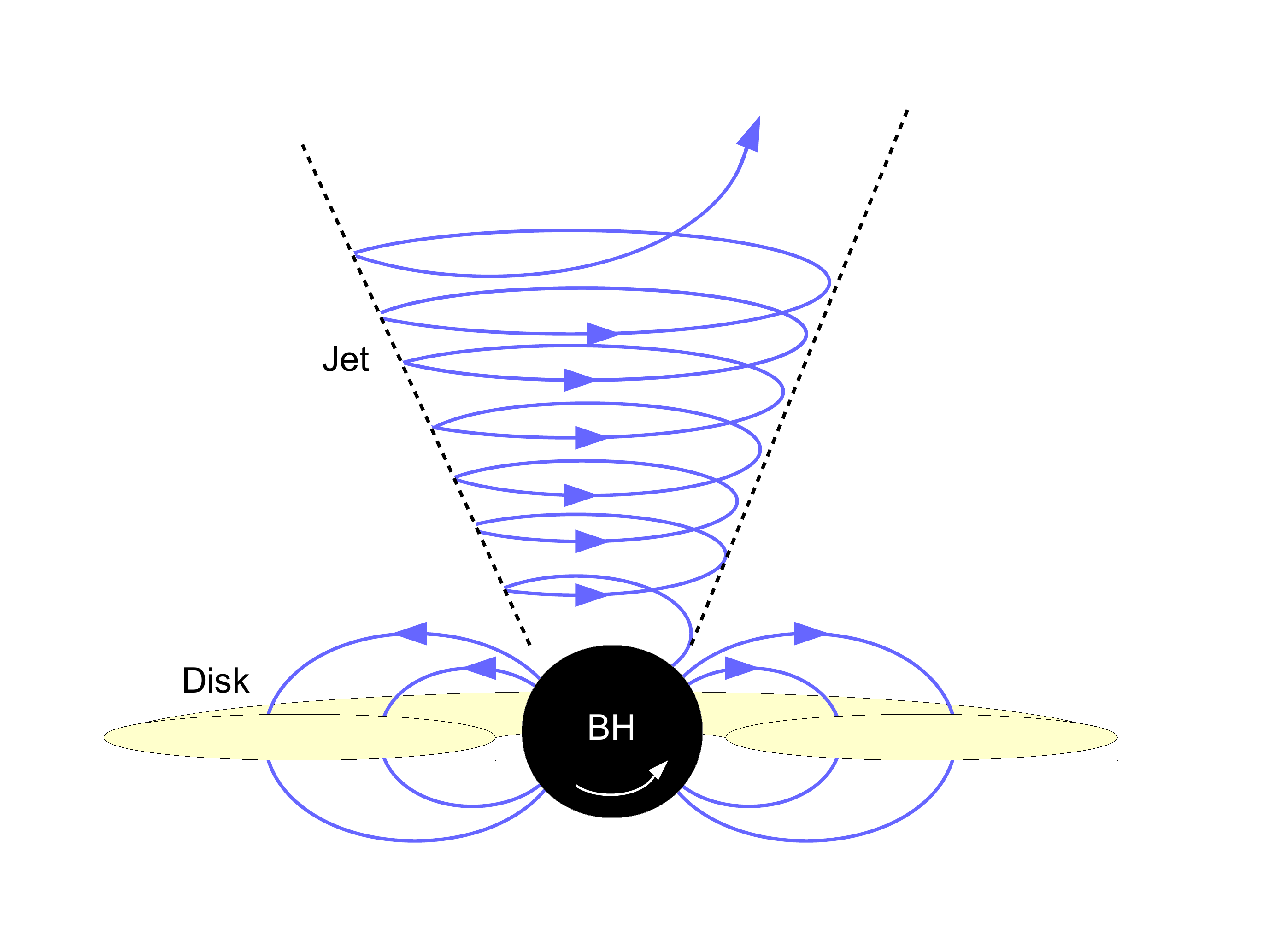}\\
\caption{Relativistic jet launching from a black hole-accretion disk system, formed as a result of a PBH-NS interaction, through the Blandford-Znajek mechanism. Magnetic field lines (blue) are anchored in the accretion disk (yellow) and penetrate the black hole's ergosphere. Spinning BH twists open field lines, leading to a jet.}
\label{fig:schematic}
\end{figure} 

Previously~\cite{Fuller:2017uyd}, we have shown that up to $\sim0.1-0.5 M_{\odot}$ of material can be ejected from a NS rotating near mass-shedding limit if all of the angular momentum from particles infalling to the growing PBH is efficiently transfered to the star's outer shells. On the other hand, it is likely that some angular momentum will be transfered to the BH instead. Here we consider that the BH acquires majority of the star's angular momentum. This will result in an increase of the BH's Kerr spin parameter $a$, which in turn decreases its innermost stable orbit (ISCO) radius $r_{\rm ISCO}$.~For a BH of mass $M_{\rm BH}$, the ISCO radius can vary from $r_{\rm ISCO} (a = 0) = 3 R_{\rm S}$ for a non-spinning Schwarzschild BH to $r_{\rm ISCO} (a = 1) = R_{\rm S}/2$ for a maximally spinning Kerr BH, with $R_{\rm S} = 2 G M_{\rm BH}$ denoting the BH's Schwarzschild radius.~Any residual material outside of $r_{\rm ISCO}$ will participate in formation of the accretion disk.~We estimate that a $\sim10^{-2}-10^{-1} M_{\odot}$ disk will form in a generic NS-PBH system (see Appendix~A). Upcoming simulations will allow to definitively verify this claim. As a base reference for further estimates we take a typical disk surrounding the resulting BH to be of
$M_{d} \sim 0.1 M_{\odot}$ size and further consider it to be accreted within $\Delta t \sim 0.1$ s.

As alluded to above, the sGRB jet engine \cite{Lee:2007js} can be powered by extracting energy from the accretion disk. For a maximally rotating BH up to  42\% of the disk's binding energy  $E_{b} = 0.42 M_{\rm d} \simeq 10^{53} (M_d / 0.1 M_{\odot})$ erg is extractable.~The two major production mechanisms are neutrino--anti-neutrino annihilation and magneto-hydrodynamic (MHD) winds (Blandford-Payne~\cite{Blandford:1982di}). 

For effective neutrino-anti-neutrino annihilation  one needs extreme conditions with temperatures reaching $T \sim 10^{11}$ K, such that the photons are completely trapped \cite{Lee:2007js}.
Unlike the standard post-merger NS and the collapsar scenarios, in our setup there is no significant source of potential energy that will result in a heated accretion disk. On the other hand, to allow for accretion rates of order $\sim 0.1 M_{\odot}$/s to produce 
sGRB a very efficient turbulent viscosity is necessary to transport the angular momentum from the accretion disk. This effective viscosity will heat up matter. The resulting heating can be estimated from the Shakura-Sunyaev $\alpha$-disk model\cite{Shakura:1972te}, with the mid-plane disk temperature being
\begin{equation}
T_c = 1.4 \times 10^4 \alpha^{-1/5} \dot{M}_{16}^{3/10} m_1^{1/4} R_{10}^{-3/4} f^{5/6}~\text{K}~,
\end{equation}
where $\alpha$ is a phenomenological $\mathcal{O}(1)$ parameter that captures the effective viscosity, $\dot{M}_{16}$ is accretion rate in units of $10^{16}$ g/s, $m_1$ is the mass of the central accreting object in units of $M_{\odot}$, $R_{10}$ is the radius of a point in a disk in units of $10^{10}$ cm and $f = [1 - (R_{\star}/R)^{1/2}]^{1/4}$, with where $R_{\star}$ is the radius where angular momentum stops being transported inwards. For accretion rate of $0.1 M_{\odot}$/s and a point at the ISCO radius $R = r_{\rm ISCO}(a=0) = 10$ km, assuming a solar mass Schwarzschild BH, we observe that the temperatures can exceed $T_c \sim 10^{12}$ K and neutrino-antineutrino annihilation could thus become effective. With neutrino luminosities of $L_{\nu} \sim 10^{52} (M_{\rm disk}/0.1 M_{\odot}) (\Delta t/1~\text{s})^{-1} $ erg/s and pair conversion efficiencies of $L_{\nu \overline{\nu}} \sim 10^{-3} L_{\nu}$ this channel could therefore in principle also contribute to sGRB production, via $\nu \overline{\nu} \rightarrow e^+e^- \rightarrow \gamma\gamma$ \cite{Lee:2007js}. 

For MHD winds, strong magnetic fields are required.
Shearing induced by differential rotation of the disk as well as instabilities~\cite{Meier:1998fw} are expected to amplify the disk's magnetic fields to magnetar-like $\sim10^{15}$ G levels (e.g.~\cite{Rezzolla:2011da}). If the magnetic field does not penetrate the BH and is confined to the disk, the disk will radiate as an electro-magnetic dipole similar to a pulsar (see model of \cite{Usov:1992zd}) with a resulting luminosity of $L_{\rm EM} \sim 10^{49} (B/10^{15}\text{G})^2(P/1\text{ms})^{-4}(R/10\text{km})^6$ erg s$^{-1}$ for a solar mass Schwarzschild BH with $R = r_{\rm ISCO}(a=0)$, sourcing sGRB from MHD winds.
 
An alternative sGRB power source is the BH itself. The resulting solar-mass BH can harbor significant rotational energy.~When the disk's magnetic field penetrates the BH, rotational energy can be extracted from the BH's ergosphere through the Blandford-Znajek (BZ) mechanism \cite{Blandford:1977ds} as depicted on Figure~\ref{fig:schematic} (right).~For a maximally spinning BH up to 29\% of its rest mass is extractable \cite{Punsly:book2001}.~The total NS angular momentum is given by $J_{\rm NS} = \Omega_{\rm NS} I_{\rm NS}$, where $I_{\rm NS} = (2/5) M_{\rm NS} R_{\rm NS}^2$ is its moment of inertia. With the resulting BH retaining most of the original NS's mass and angular momentum, its Kerr parameter will be $a = J_{\rm BH}/G M_{\rm BH}^2 \simeq J_{\rm NS}/ G M_{\rm NS}^2 = 0.53 (P/ 1\text{ms})^{-1}$.~The available energy for extraction is the reducible BH mass, given by
\begin{equation}
M_{\rm red} = M_{\rm BH}\Big[1 - \Big(\dfrac{f(a)}{2}\Big)^{1/2} \Big] \simeq 10^{53}~\text{erg}~,
\end{equation}
where $f(a) = (1 + \sqrt{1 - a^2})$.
The frame-dragging BH angular velocity at the outer event horizon $R_{+} = f(a) R_{\rm S}/2$ is 
$\Omega_H = a/2 R_{+} \simeq 2 \times 10^4~\text{rad}~\text{s}^{-1}$.
For a uniform magnetic field $B_0$ aligned with rotation axis of the BH in vacuum (i.e.~Wald solution \cite{Wald:1974np}) the flux through BH's hemisphere is given by \cite{Punsly:book2001}
\begin{equation}
\Phi_{\rm BH} = \pi R_{+}^2 B_0 \Big[ 1 - a^4 \Big( \dfrac{R_{\rm S}}{2 R_{+}}\Big)^4\Big]~.
\end{equation}
The resulting BZ luminosity is thus \cite{Blandford:1977ds,Tchekhovskoy:2009ba} 
\begin{equation} \label{eq:lumBZ}
L_{\rm BZ} \simeq \dfrac{\kappa}{4\pi} \Omega_{\rm H}^2 \Phi_{\rm BH}^2 = 6 \times 10^{47} \Big(\dfrac{B_0}{10^{15}\text{G}}\Big)^2 ~\text{erg}~\text{s}^{-1}~,
\end{equation}
where constant $\kappa \simeq 0.05$ weakly depends on the magnetic field geometry.~The above approximation is accurate up to $a \sim 0.95$ \cite{Tchekhovskoy:2009ba}, which covers our region of interest. Hence, the total energy output of the process is $E_{\rm BZ} = L_{\rm BZ} \Delta t \sim  10^{47}$ erg. 
It has been also recently speculated  \cite{Lyutikov:2011tk} (however, see \cite{Lehner:2011aa}) that magnetosphere could allow BHs to retain ``magnetic hair'' for a long time after the magnetized disk has been accreted, enabling BHs to source the BZ jets themselves, resulting in another production mechanism.

We note that magnetic fields of $\sim10^{16}$ G strength can already be found in magnetars \cite{Duncan:1992hi,Kaspi:2017fwg}, which have also been proposed as sources of GRBs. While the magnetar population could be significant, magnetic braking and rapid radiative spin-down renders their contribution to our setup insignificant.

{\bf Microquasars -- } Microquasars \cite{Mirabel:1994rb,Sams:1996} are X-ray binaries (XRBs) with accreting stellar-mass compact objects (NS or BH) that exhibit relativistic jets.~Their broad emission spectrum \cite{Romero:2016hjn} spans many decades in energy, ranging from eV up to and above TeV. PBHs can also form MQs. The simplest possibility is companion capture by a stellar-mass PBH. However, such PBHs are already strongly constrained if they are to constitute DM. Here we envision a more subtle scenario, realized when a small PBH consumes a binary star (NS or WD) and transforms it into a jet-emitting BH.

We first consider the case of forming a BH MQ from a NS. Even before turning into BHs, neutron star XRBs are likely already emitting relativistic jets.~Still, the resulting behavioral change is observable.~One discriminating signature between them is emission of the keV thermal X-ray component originating from the plasma-star surface interactions present in the spectrum of XRBs with NSs but not BHs, which lack star surface~\cite{Meier:2012book}.~In principle, black hole MQs could also be significantly more efficient emitters. The associated
BZ mechanism has the potential to extract energy at a rate higher than the BH's energy inflow, exceeding 100\% efficiency, as occurs for magnetically-arrested disks \cite{Tchekhovskoy:2015}. 
Observational evidence for BZ mechanism in stellar-mass BH XRBs is controversial \cite{Fender:2010tk,Narayan:2011eb}.~For further comparison between NS and BH jets see \cite{Migliari:2005hv}. 

Interacting white dwarfs also possess accretion disks and can emit jets.~While there have been proposals to model them after MQs, as ``nano-quasars'' \cite{Zamanov:2002bw}, their jets are non-relativistic ($v \sim 10^{-2}$) and luminosity is strongly sub-Eddington ($L \lesssim 10^{-2} L_{\rm Edd}$).~This picture can dramatically change if a PBH transforms an accreting WD into a BH. The resulting BH mass will be similar to the original WD, $M_{\rm WD} \sim 1 M_{\odot}$.~The accreting distance, however, will drastically decreases from the WD radius of $R_{\rm WD} \simeq 10^{3}-10^{4}$ km to the BH ISCO radius of $r_{\rm ISCO} < 10$ km.~For a constant accretion rate $\dot{M}$ the respective luminosity is given by $L_{\rm acc} = \epsilon \dot{M}$, where $\epsilon = G M / R$ is the efficiency parameter \cite{Meier:2012book}.~Hence, the accretion efficiency of the system increases from $\sim10^{-4}$ for the WD to $\sim10^{-1}$ for the BH, which could enable luminosity to reach near-Eddington levels as desired for microquasar jets \cite{Romero:2016hjn}.
Since a typical WD rotates with a period of  hours-to-days \cite{Kawaler:2003sr}, the spin parameter of the (post-WD) resulting BH is negligible.~Thus, activation of a jet through BZ is not effective and other mechanisms will need to be invoked.~For a Blandford-Payne -type setup the jet luminosity will depend on the accretion as $L_{\rm jet} = (1/2) q_{\rm jet} L_{\rm acc}$, where $q_{\rm jet} < 1$.~The decrease in the compact object radius will also result in an increase of the temperature in the surrounding accretion disk.~For a thin Shakura-Sunyaev disk~\cite{Shakura:1972te} the maximum temperature scales with the disk's inner radius $r_0$ as $T_{\rm d}^{\rm max} \propto (\dot{M} M / r_0^3)^{1/4}$.~Provided a constant accretion rate, active galactic nuclei-MQ jets are known to obey general scaling relations with size of the central black hole~\cite{Sams:1996,Rees:2004tz}.~Hence, the considered solar-mass BH MQs could in principle produce similar output as typical MQs with BHs of some-what higher mass.

{\bf Positrons -- } Both GRBs as well MQ jets can accelerate particles to high energies.~First we comment on GRBs.~One way to ensure that jet possesses a high relativistic factor $\Gamma$ is by invoking BZ mechanism, whose high magnetic fields will prevent jet contamination by protons.~GRBs are expected to be sources of copious 
electron-positron pair production at the MeV energies~\cite{Thompson:1999vc,Dermer:2000zs}. A sizable population of GeV-TeV positrons can appear from re-scattering of TeV photons from the initial burst on a low energy target photon field, resulting in $\gamma\gamma \rightarrow e^+e^-$. Such field can be provided by eV photons from the optical flash afterglow associated with the GRB in question \cite{Waxman:1999rm}. Assuming typical GRB energetics, the resulting total pair energy will be of similar magnitude as the GRB photons, carrying $\sim10^{50}$ erg and with a spectral index $\sim2$ \cite{Ioka:2008cv}.

For MQs, a variety of jet models exist~\cite{Romero:2016hjn}. 
Their jets include several contributions, including  the corona, the accretion disk and  the companion star fields, resulting in a complicated multi-component spectrum. A major production channel (e.g. see lepto-hadronic models of \cite{Vila:2011aa}) for energetic GeV-TeV positrons is $p \gamma \rightarrow \pi^+ n, \pi^0 p$ followed by pion decays to leptons.~Pair-production $p \gamma \rightarrow p e^+e^-$ can also significantly contribute, depending on the details of the target photon field \cite{Hu:2009bc}. The resulting emission is well modeled by a broken power-law with a spectral index of $\sim1.5-2.5$~\cite{Gupta:2014hua}. Energetic neutrinos produced from MQ muon and pion decays have also been suggested as signal sources for neutrino telescopes~\cite{Bednarek:2005gf}.

\begin{figure}[tb]
\centering
\hspace{-2em}
\vspace{-1.5em}
\includegraphics[trim = 0.0mm .0mm 20mm 0mm, clip,width = 3.5in]{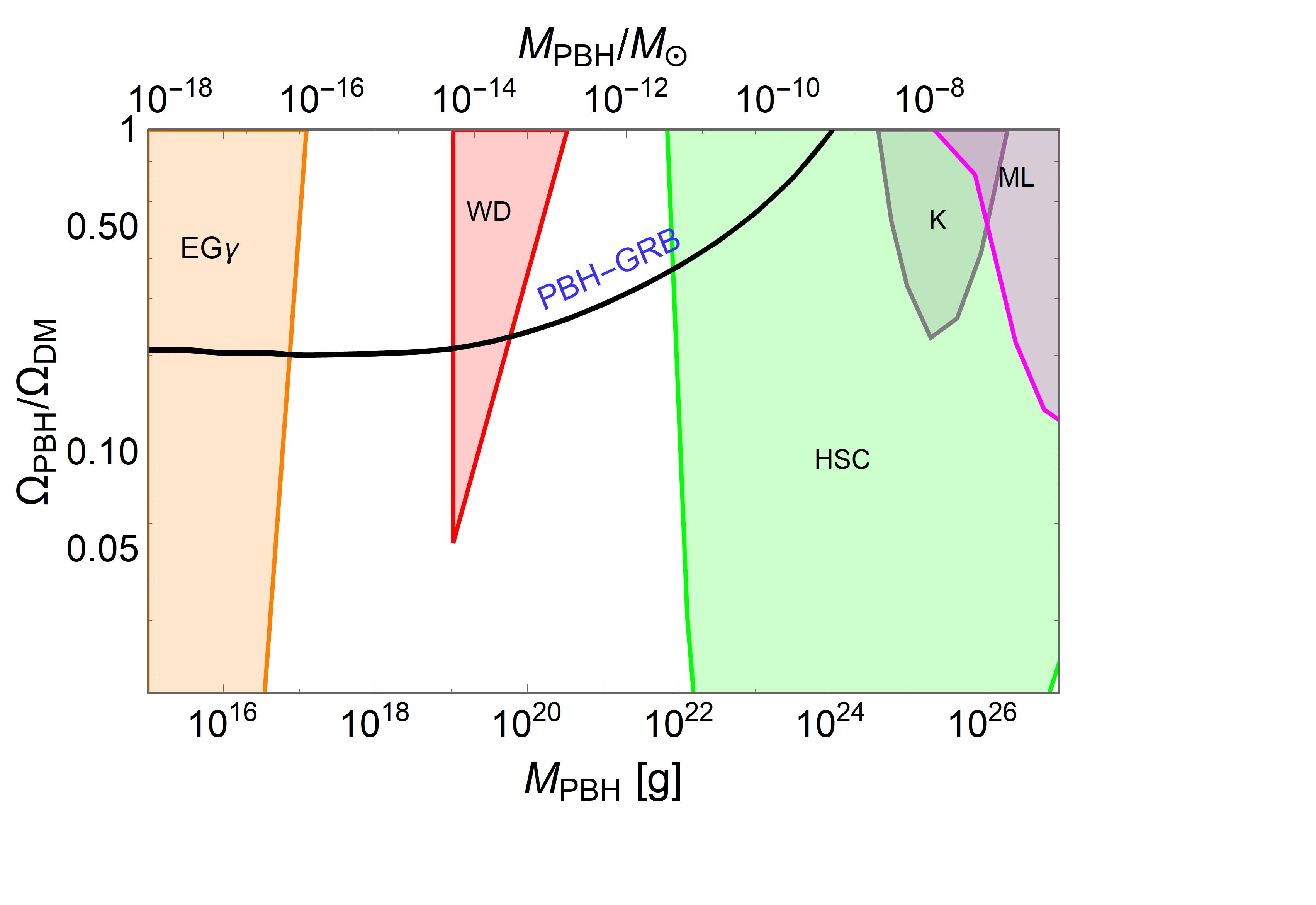}\\
\caption{
Allowed parameter space for PBH-GRBs to significantly contribute the observed positron excess at energies $\gtrsim 10$ GeV.~The black curve represents the signal rate calculated with the most optimistic input parameter choice.~The allowed parameter space for PBH-GRB positrons contributing to the excess lies above this line.
Shaded regions denote parameter space constrained by extragalactic $\gamma$-rays from BH  evaporation (EG$\gamma$)~\cite{Carr:2009jm}, white dwarf abundance (WD)~\cite{Graham:2015apa}, Kepler star milli/micro- lensing (K)~\cite{Griest:2013aaa}, Subaru HSC micro-lensing (HSC)~\cite{Niikura:2017zjd} and MACHO/EROS/OGLE micro-lensing (ML)~\cite{Tisserand:2006zx}. }
\label{fig:PBHGRB}
\end{figure}

The emitted positrons diffuse, with evolution characterized by the diffusion equation.~They will lose energy primarily through synchrotron and inverse Compton scattering. The associated energy losses can be described by $dE/dt = - \beta E^2$, where $\beta \simeq 1.6 \times 10^{-16}$ GeV$^{-1}$ s$^{-1}$~\cite{Porter:2005qx}. The diffusion time to lose half of initial energy $E_0$ is $\tau_d = 1/ \beta E_0$, which for 100 GeV positrons translates to $\tau_d (100~\text{GeV}) \simeq 2.1 \times 10^6$ yrs.~The diffusion radius is given by $r_d \simeq 2 \sqrt{K \tau_d}$. Following \cite{Malyshev:2009tw}, we adopt the diffusion coefficient of $K(E) = K_0 (1 + E/1~\text{GeV})^{\delta}$, with $K_0 = 3 \times 10^{28}$ cm$^{2}$ s$^{-1}$ and $\delta = 0.4$. Hence, for 100 GeV positrons $r_d (100~\text{GeV}) \simeq 2.2$ kpc. 

{\bf Signal rate -- }~For a burst point source with a power-law spectrum, such as a GRB, the diffusion equation can be solved analytically (e.g.~\cite{Aharonian:1995zz}). In \cite{Gupta:2014hua} the authors have fitted the observed positron excess to the diffused MQ spectrum, assuming a uniform Galactic distribution of XRBs that are currently emitting. They found that $\mathcal{O}(10^{2}-10^{3})$ MQs with a typical Eddington luminosity of $\sim 10^{38}$ erg s$^{-1}$ and emitting $\sim10^{32-34}$ erg s$^{-1}$ positron flux can account for the excess.
In \cite{Ioka:2008cv} it was shown that a single GRB burst that occurred at a distance $\sim$1 kpc away within the diffusion time, or conversely a single MQ that has been continuously emitting for such duration, can both fit well the positron excess.

MQs or GRBs formed as a result of PBH-interactions will contribute to the positron excess in an analogous manner. For PBHs constituting DM, the strongest signal originates from locations where DM as well as compact star densities are the highest, which occurs in the Galactic Center.~However, since the density of diffusing particles follows $(1/r)\text{erfc}(r/r_d)$, with $\text{erfc}$ being the error function, contributions from outside the diffusion radius are highly suppressed \cite{Atoyan:1995}. Thus, the excess is dominated by local sources. The fraction of dark matter in the form of PBHs that is required to address $N_{\rm sig}$ positron signal sources necessary for the excess can be estimated as
\begin{equation} \label{eq:fitform}
\Big(\dfrac{\Omega_{\rm PBH}}{\Omega_{\rm DM}}\Big) = \dfrac{N_{\rm sig}}{F_0  \, \tau_d \, N_{\rm st}}~,
\end{equation}
where $N_{\rm st} = n_{\rm star} V_d$ is the relevant compact star population, with $n_{\rm st}$ and $V_d = (4/3) \pi r_d^3$ denoting the local population density of stars and the diffusion volume, respectively. 
The density of WD interacting binaries, taken to be cataclysmic variables, is predicted to be $10^{-5} - 10^{-4}$ pc$^{-3}$ \cite{Ak:2007ge}. The neighborhood NS density is given by $(1-5) \times 10^{-4}$ pc$^{-3}$ \cite{Sartore:2009wn}.
The system interaction timescales are set by the diffusion time $\tau_{\rm d}$.

For PBH-induced MQs, contributions to the positron excess can come either from WDs or NSs. Since PBH-WD capture rate is several orders below that of NS, these interactions are sub-dominant to PBH-NS induced GRBs or MQs.
On the other hand, NS XRBs will already be emitting and contributing positrons even prior to interactions with PBHs. While the pre- and the post-PBH capture jet emission from these systems is not expected to be identical, we shall not discuss PBH-NS MQ systems further, focusing solely on PBH-NS GRBs below. We note, that continuous MQ jets can also contribute to the observed proton flux \cite{Aguilar:2016kjl}. However, because of low WD-PBH capture rates, this is not significant.

In order to contribute to the excess through PBH-induced GRBs with a high likelihood, we require formation of at least one local source from a millisecond pulsar on the relevant timescales~\cite{Ioka:2008cv}. The population of millisecond pulsars and pulsars is expected to be comparable~\cite{Lorimer:2008se}.~Since millisecond pulsars originate from binaries, we double the capture rate to approximate their stronger gravitational potential.~For simplicity, we implicitly assume that all PBH-NS systems will result in a sGRB.

To fit Eq.~\eqref{eq:fitform} we scan over a broad range of astrophysical input parameters. The local DM density is varied as $0.2~\text{GeV}~\text{cm}^{-3} \lesssim \rho_{\text{DM}} \lesssim 0.4$ GeV cm$^{-3}$ \cite{Weber:2009pt}.
The DM velocity dispersion values are considered to lie in the 
$50~\text{km/s} < \overline{v} < 200~\text{km/s}$ range. Here, the lower $\overline{v}$ limit corresponds to a possible DM disk within the halo~\cite{Read:2008fh,Read:2009iv}, while the upper limit corresponds to the Navarro-Frenk-White DM density profile without adiabatic contraction~\cite{Kaplinghat:2013xca}.~The effects of natal pulsar kicks are included, modifying the capture rate $F_0$ according to the method outlined in \cite{Fuller:2017uyd}. The pulsar velocity dispersion is varied between 48 km/s \cite{Cordes:1997my} and 80 km/s \cite{Lyne:1998}. We also consider uncertainty on the diffusion parameter, which at least in some regions could be smaller by $\sim10^2$ than typical values, as indicated by the observations from HAWC \cite{Abeysekara:2017old}. In more detail, since $r_d \propto \sqrt{K_0}$, assuming such a coefficient throughout all regions will result in decrease of $r_d$ by an order of magnitude. Hence, the corresponding diffusion volume, along with the number of potential star sources $N_{st}$, will be reduced by a factor of $\sim 10^3$. Thus, the required PBH DM abundance will also respectively increase, through Eq.~\eqref{eq:fitform}.

We have considered PBH-NS GRB contributions to the excess in two positron energy regimes, high ($\gtrsim 100$ GeV) and low ($\gtrsim 10$ GeV).
In the high energy regime we estimate that sizable PBH-NS GRB contributions are improbable, throughout the scanned parameter space. On the other hand, PBH-NS GRBs can contribute significantly in the low energy regime, where the diffusion radius and the diffusion time increase to $r_d \simeq 4.5$ kpc and $\tau_d \simeq 2.1 \times 10^7$ yrs, respectively. 
 
The fit results are displayed on Fig.~\ref{fig:PBHGRB}, along with experimental constraints. The enclosed region above the black curve displays the parameter space for PBH-GRBs to contribute significantly the positron excess at energies $\gtrsim 10$ GeV.~We note that, in the mass ranges of $10^{17} - 10^{19}$ g and $10^{20}-10^{23}$~g, PBHs can account for all of the dark matter. 
In the same mass range, one could use the stability of neutron stars to constrain PBHs if globular clusters contained $\gtrsim 10^3$ times the average dark matter density~\cite{Capela:2013yf}. However,
observations of globular clusters show no evidence of  dark matter content in such systems, resulting in upper bounds three order of magnitude below the levels needed to allow for meaningful constraints~\cite{Bradford:2011aq,Sollima:2011nb}.~Further, potential GRB femtolensing constraints around $\sim10^{18}$ g PBH mass range are also ineffective, due to extended nature of the sources as well as wave optics effects~\cite{Katz:2018zrn}. Large astrophysical uncertainties could vary the results by several orders.~Considered timescales assume that a stellar-mass BH will form rapidly after PBH has been captured. For $M_{\rm PBH} \lesssim 10^{20}$ g the time for PBH to settle inside the star will exceed the interaction time set by $\tau_{\rm diff}$.~On the other hand, lower mass PBHs that have been captured earlier will then contribute at the same rate.~Such behavior occurs until the settle time will reach the Galactic lifetime of $\sim 10^{10}$ yrs, which will happen for $M_{\rm PBH} \lesssim 10^{18}$ g.~At lower masses, however, the capture rate and settle time are not well understood.

In our analysis we have assumed a monochromatic
PBH mass function, allowing for a general study. While
typically the PBH formation models predict an extended
mass function, the details are highly model-dependent.
A procedure for implementing an extended mass function,
corresponding to a particular formation model, is
outlined in \cite{Carr:2017jsz}.

We further comment on the sGRB rates from PBHs compared to those from binary neutron stars. Observation of GRB 170817A accompanying the gravitational signal GW170817 from NS-NS merger have confirmed compact object mergers as a class of sGRB progenitors~\cite{Monitor:2017mdv}. Unlike GWs, it is hard to establish a clear relationship between sGRB distance and brightness, with varying signal luminosity as well as beaming angle affecting signal detection. Assuming that Galactic NS-NS mergers as well as PBH-induced events will result in observed sGRBs and that NS-NS are the main sGRB progenitors, the Milky Way merger rate of $R_{\rm MW-NS} \sim 20$ Myr$^{-1}$ \cite{Chruslinska:2017odi} can also be in principle interpreted as a possible constraint on PBH-NS encounters, subject to many assumptions and uncertainties.

{\bf Conclusions -- }
We have proposed several scenarios how PBHs can lead to sGRBs and MQs through interactions with compact stars, providing novel astrophysical signatures. While sGRBs typically originate from binary mergers, the resulting PBH-sGRBs will not have accompanying merger GWs. On the other hand, PBH-induced MQs will show a sudden permanent change in the emission spectrum of the system. This allows to distinguish these systems from the usual astrophysical ones.

Positrons accelerated in relativistic jets of locally-formed PBH-induced sources can contribute to the positron excess, as observed by several experiments.~The presented proposal thus
combines beneficial features of astrophysical sources with dark matter, which are typically disjoint in other proposals related to the excess. The resulting relevant parameter space for PBHs resides in the open window where they can constitute all of DM and is consistent with the range where PBHs can also shed light on other astronomy puzzles \cite{Fuller:2017uyd}. Further confirmation of standard astrophysical sources as the origin of the positron excess (e.g. pulsars \cite{Hooper:2017gtd}) will allow to constrain PBHs as DM in a certain region of the parameter space.

{\bf Acknowledgments -- }
We thank Alexander Kusenko for originally suggesting the problem investigated in this work as well as fruitful discussions throughout the project. We further thank Ping-Kai Hu, Yoshiyuki Inoue and Edward Wright for comments. This work was
supported by the U.S. Department of Energy Grant No. DE-SC0009937.
 \(\)
 
 \appendix
 \section{Appendix A: Accretion Disk Formation} 

Accretion disk formation in a collapsing star system can be analyzed by estimating the amount of material residing outside the ISCO radius of the resulting black hole~\cite{Shapiro:1983du,Shapiro:2004zx}. 

Rotating milli-second pulsars spin near mass-shedding limit and can be described analytically by an extended Roche spheroid model \cite{Shapiro:1983du,Shapiro:2004zx}.~The equatorial radius $R_{\rm eq}$ is stretched and exceeds the polar radius $R_p$, with $R_{\rm eq} = (3/2) R_p$.~Modeling the rapidly rotating star as a polytrope of index $n$, the spherical-coordinate density of the extended envelope is given by~\cite{Shapiro:2004zx}
\begin{align} 
\rho(r, \theta) = ~& \dfrac{\xi_1^{3 - n} (\xi_1^2 |\theta^{\prime}(\xi_1)|)^{n-1}}{4 \pi} \dfrac{M}{R_p^3} \notag\\
& \times \Big(\dfrac{R_p}{r} - 1 + \dfrac{4}{27} \dfrac{r^2}{R_p^2} \sin^2(\theta)\Big)^n~,
\end{align}
where $\theta(\xi)$ is a solution
to the Lane-Emden equation and $\xi$ denotes dimensionless radius, with $\xi_1$ corresponding to the star's radius extent. The surface boundary is located along the $\rho = 0$ curve and can be described as~\cite{Shapiro:2002kk}
\begin{equation} \label{eq:starsurf}
r(\theta) = \dfrac{ 3 R_p \sin(\theta/3)}{\sin(\theta)}~.
\end{equation}

We estimate the amount of angular momentum and mass located within the resulting BH that has consumed the star from within up to the polar radius $R_p$ through spherical accretion.~At that point a wholesome NS is no longer present.~With rotational symmetry in play, the ratio of the resulting BH mass to the original NS mass can be found from
\begin{equation}
\dfrac{M_{\rm BH}}{M_{\rm NS}} = \dfrac{\int_{0}^{\pi/2} \int_{0}^{R_p} \rho(r, \theta) \, r^2 \sin(\theta) dr d\theta}{\int_{0}^{\pi/2} \int_{0}^{r(\theta)} \rho(r, \theta) \, r^2 \sin(\theta) dr d\theta}~.
\end{equation}
To make further progress, we need to assume a NS EoS or in our context a corresponding polytropic index. As discussed in the text, recent observations favor stiffer (lower $n$) NS profile.~Since analytic solutions for the Lane-Emden equation are only known for $n = 0, 1, 5$, we take $n = 1$ as a NS description.~We thus find that the mass ratio of the resulting BH compared to the original NS is $M_{\rm BH} / M_{\rm NS} \simeq 0.9$, which for $M_{\rm NS} = 1.5 M_{\odot}$ results in $M_{\rm BH} \simeq 1.4 M_{\odot}$.

As before, we take that all of the angular momentum from the in-falling particles is not transferred outside but is acquired by the BH instead.~Hence, the resulting BH angular momentum is $J_{\rm BH} = (2/5) \Omega_{\rm NS} M_{\rm BH} R_p^2$, yielding a spin parameter of $a \simeq 0.6$.
The corresponding ISCO radius $r_{\rm ISCO}(a = 0.6) \simeq 4 G M_{\rm BH}$ is found from \cite{Shapiro:1983du}
\begin{equation}
r_{\rm ISCO} = G M_{\rm BH} \Big[3 + Z_2 - \sqrt{(3-Z_1)(3+Z_1+2 Z_2)}\Big]~,
\end{equation}
where
\begin{equation}
Z_1 = 1 + (1 - a^2)^{1/3} [(1+a)^{1/3} + (1-a)^{1/3}]
\end{equation}
and $Z_2 = (3 a^2  + Z_1^2)^{1/2}$.

The final equatorial radius after BH has consumed the star up to $R_p$ is given by 
$R_{\rm eq}^f = R_{\rm eq} - R_{\rm p} + R_{+}$, where $R_{+}$ is the outer Kerr BH horizon as before.
The ratio of the resulting disk mass $\Delta M$ to the original star's mass $M_{\rm NS}$ can then be found through summation of the remaining cylindrical mass shells outside the ISCO radius as proposed in \cite{Shapiro:2004zx}: 
\begin{align}
& \dfrac{\Delta M}{M_{\rm NS}} =  \xi_1^{3 - n} (\xi_1^2 |\theta^{\prime}(\xi_1)|)^{n-1}  \\
& \times  \int_{r_{\rm ISCO}^{\prime}}^{R_{\rm eq}^{f^{\prime}}} \int_{0}^{z^{\prime}(r)}
r  \Big(\dfrac{1}{\sqrt{r^2 + z^2}} - 1 + \dfrac{4}{27}r^2 \Big)^{n} dz dr~, \notag
\end{align}
where the prime on top of quantities in limits of integration denotes normalization to $R_p$ as $R_{\rm eq}^{f^{\prime}} = R_{\rm eq}^{f}/R_p$, $r_{\rm ISCO}^{\prime} = r_{\rm ISCO} / R_p$ and $z^{\prime}(r) = z(r)/R_p$. Here, $z(r)$ describes the height of a cylindrical shell and can be in principle obtained by changing coordinates and then inverting Eq.~\eqref{eq:starsurf}. We approximate the height by performing the $dz$ integration up to $R_{+}^{\prime}/2$, with $R_{+}^{\prime} = R_{+}/R_p$.
We thus obtain $\Delta M / M_{\rm NS} \simeq 0.1$, with the resulting disk mass of $M_{\rm disk} \simeq 0.1 M_{\odot}$. We stress that our analytic estimates are rather crude and do not include relativistic effects, but we hope that they will serve as additional motivation for upcoming simulations of PBH-NS systems to analyze such issues in higher detail.

\bibliography{pbhpos}
\end{document}